\documentclass[11pt]{amsart}
\usepackage{latexsym,amssymb,amsmath,amscd,amsthm}
\usepackage{amsfonts}
\numberwithin{equation}{section}
\theoremstyle{remark}

\newcommand{\bq}{\begin{equation}}
\newcommand{\bea}{\begin{array}}
\newcommand{\eea}{\end{array}}

\newcommand{\ga}{\alpha}

\newcommand{\gD}{\Delta}
\newcommand{\gl}{\lambda}

\newcommand{\gb}{\beta}

\newcommand{\mf}{\mathfrak}

\newcommand{\mc}{\mathcal}

\newcommand{\dg}{\dagger}

\newcommand{\go}{\omega}

\newcommand{\gG}{\Gamma}

\newcommand{\gs}{\sigma}

\newcommand{\gag}{\gamma}
\newcommand{\gd}{\delta}
\newcommand{\pp}{\partial}

\newcommand{\na}{\nabla}

\newcommand{\bl}{\blacklozenge}
\newcommand{\bs}{\blacksquare}

\newcommand{{\DDD}}{D\!\!\!\!\!\!-}

\title{ON A DEFORMED QUANTUM POTENTIAL}

\author{Robert Carroll\\University of Illinois, Urbana, IL 61801}

\date{November, 2012\thanks{email: rcarroll@math.uiuc.edu}}

\begin{document}

\bibliographystyle{plain}

\maketitle

\section{BACKGROUND}
\renewcommand{\theequation}{1.\arabic{equation}}
\setcounter{equation}{0}

In the rather incomplete survey paper \cite{crrr} among other things we overlooked the fact that the Schr\"odinger SE) for a modified
Riemann-Liouvill (MRL) calculus in \cite{jmre} actually gives rise to a ``heuristic" quantum potential
(QP) directly as follows.  The SE for MRL from \cite{jmre}, 41 (2009), 1590-1604 involves
\bq\label{1.1}
i\hbar\psi_t^{\ga}=-\frac{\hbar^2}{2m}\rho^2(\ga)(xt)^{2(\ga-1)}\psi_{xx}+V\psi
\end{equation}
where $\rho(\ga)=\ga![(1-\ga)!]^2$ and the momentum $p_{\ga}\sim m\dot{x}^{(\ga)}=mu_{\ga}$
where 
\bq\label{1.2}
u_{\ga}=\frac{(dx)^{\ga}}{dt}=\ga!\left(\frac{d^{\ga}t}{dx}\right)^{-1}
\end{equation}
Thus $\dot{x}^{\ga}$ is defined by the $\ga$-derivative of time with respect to space
(recall $\ga!d^{\ga}t=dt$) and consequently
\bq\label{1.3}
u_{\ga}=\rho(\ga)(xt)^{(\ga-1)}x^{\ga}(t)
\end{equation}
(cf. \cite{crrr,jmre}).  Then when considering $p_{\ga}^2$ in the Hamiltonian one simply
uses $u_{\ga}^2$.  It is not immediately clear why this is a better determination of velocity
than e.g. $({\bf 1A})\,\,v_{\ga}(t)=d^{\ga}x/(dt)^{\ga}=\ga![dx/(dt)^{\ga}]$.  Also, given a 
classical time variation one could also imagine a SE ($D^{2\ga}=D^{\ga}D^{\ga}$) ($D^{2\ga}=
D^{\ga}D^{\ga}$)
\bq\label{1.4}
i\hbar\psi_t=-\frac{\hbar^2}{2m}D^{2\ga}\psi+V\psi
\end{equation}
arising via $(\bullet)\,\,\psi=Rexp[iS/\hbar]$ or $\psi=RE_{\ga}(z^{\ga})$ (where $E_{\ga}$ is the Mittag-
Leffler function $({\bf 1B})\,\,E_{\ga}=E_{\ga}(z^{\ga})=\sum[z^{(\ga k)}/(\ga k)!$ with $D^{\ga}E_{\ga}=E_{\ga}$).  In the latter situation one would have a possible quantum potential (QP) for (1.4) of the form
\bq\label{1.5}
{\mf Q}_{\ga}=-\frac{\hbar^2}{2m}\frac{D^{2\ga}R}{R}
\end{equation}
and in fact it might be suggested that (1.1), with a classical time derivative, might have a QP of the form
\bq\label{1.6}
\hat{{\mf Q}}=-\frac{\hbar^2}{2m}\rho^2(\ga)(xt)^{2(\ga-1)}\frac{R_{xx}}{R}
\end{equation}
(cf. \cite{crrr}).
Given the connection between q-deformed calculi and fractional calculi (see e.g. \cite{c002,crrr,hrmn}), and the general interest in q-deformed physics, we want to explore the
notion of a q-deformed QP in the q-deformed and fractional calculations.  However we recall
that the connection between a QP, osmotic velocity, and information theory provides the QP with its important
intrinsic meaning and it is not clear whether this arises in the fractional or deformed situations.
In particular it may not be realistic to introduce the form $iS/\hbar$ via $(\bullet)$ in the
fractional theory.  Of course if the resulting equations have solutions they may mean something 
or they may suggest a more profitable direction.
\\[3mm]\indent
{\bf REMARK 1.1.}
The presence of the time variable in in (1.6) shows that time makes its presence felt more
strongly in the fractional or deformed context (via $u_{\ga}$ for example).  Other work in
\cite{hrmn,lask,trsv} suggests memory effects in fractal situations, etc.  In particular
q-deformed and fractional contests are related (cf. \cite{hrmn}) and given the uibiquity now of
q-entropy, q-Fisher information, q-statistics, etc. it may be that the QP plays a structural role
in the fractal world.  $\bs$

\section{Q-DEFORMED AND FRACTIONAL CALCULUS}
\renewcommand{\theequation}{2.\arabic{equation}}
\setcounter{equation}{0}

We extract here first from \cite{hrmn} (1007.1084) and refer also to other citations in \cite{hrmn}
for more information (especially \cite{bnds}).  It can be shown that the concept of q-deformed Lie algebras and the methods
developed in fractional calculus are closely related and may be combined leading to a new
class of fractional q-deformed Lie algebras (cf. also \cite{c002,kzch,klsg} for q-calculus).  In order now
to describe a  deformed Lie algebra we introduce a parameter q and define a mapping
\bq\label{2.1}
[x]_q=\frac{q^x-q^{-x}}{q-q^{-1}};\,\,lim_{q\to 1}[x]_q=x;\,\,[0]_q=0
\end{equation}
\bq\label{2.2}
D_x^qf(x)=\frac{f(qx)-f(q^{-1}x)}{(q-q^{-1})x};\,\,D_x^qx^n=[n]_qx^{n-1}
\end{equation}
Alternate constructions are indicated below (cf. \cite{bnds,c002,klsg}).  
As an example of a q-deformed Lie algebra one looks at the harmonic oscillator
in \cite{hrmn}.
The creation and annihilation operators $a^{\dg},\,\,a$ and the number operator n
generate the algebra
$({\bf 2A})\,\,
[N,a^{\dg}]=a^{\dg};\,\,[N,a]=-a;\,\,aa^{\dg}-q^{\pm 1}a^{\dg}a=q^{\mp N}$.
Via (2.1) an alternative definition for ({\bf 2A}) is given via $(\bullet\bullet\bullet)\,\,a^{\dg}a=[N]_q$ with
$aa^{\dg}=[N+1]_q$.  One defines a vacuum state with $a|0>=0$ and the action of the operators $\{a,a^{\dg},N\}$ on the basis $|n>$ of a Fock space is given by
\bq\label{2.4}
N|n>=n|n>;\,\,a^{\dg}|n>=\sqrt{[n+1]_q}\,|n+1>;\,\,a|n>=\sqrt{[n]_q}|n-1>
\end{equation}
The Hamiltonian of the q-deformed harmonic oscillator and its eigenvectors are defined via
\bq\label{2.5}
H=\frac{\hbar\go}{2}(aa^{\dg}+a^{\dg}a);\,\,E^q(n)=\frac{\hbar\go}{2}([n]_q+[n+1]_q)
\end{equation}
In \cite{hrmn} various fractional derivatives are recalled and in particular one
mentions the Caputo derivative
\bq\label{2.6}
D_x^{\ga}=\left\{\begin{array}{cc}
\frac{1}{\gG(1-\ga)}\int_0^xdx(x-\xi)^{-\ga}\pp_{\xi}f(\xi) & 0\leq \ga<1\\
\frac{1}{\gG(2-\ga)}\int_0^xd\xi(x-\xi)^{1-\ga}(\pp^2f(\xi)/\pp \xi^2) & 1\leq \ga<2
\end{array}\right.
\end{equation}
Then for $x^{n\ga}$ 
\bq\label{2.7}
D_x^{\ga}x^{n\ga}=\left\{\begin{array}{cc}
\frac{\gG(1+n\ga)x^{(n-1)\ga}}{\gG(1+(n-1)\ga)} & n>0\\
0 & n=0
\end{array}\right.
\end{equation}
The fractional derivative parameter $\ga$ can be interpreted as a deformation
parameter via $|n>=x^{n\ga}$ and
\bq\label{2.8}
[n]_{\ga}|n>=\left\{\begin{array}{cc}
\frac{\gG(1+n\ga)}{\gG(1+(n-1)\ga)}|n> & n>0\\
0 & n=0
\end{array}\right.;\,\,lim_{\ga\to 1}[n]_{\ga}=n
\end{equation}
Then via ({\bf 2A})) the standard q-numbers can be defined more or less heuristically and
there are different possibilities.  On the other hand the q-deformation based on a
fractional calculus $\ga$ is uniquely determined once a set of basis vectors is given
and the harmonic oscillator can be used as an illustration.  This means that information about
q-entropy or q-information can be connected to a $\ga-$fractional background.
\\[3mm]\indent
In order to present a more ``standard" picture we go to \cite{hrmn} for a free particle SE
(cf. also \cite{lask}).  Thus replace $x$ and $p$ by $\hat{x}$ and $\hat{p}$ to get into QM and write
$({\bf 2B})\,\,
\hat{X}f(x)=xf(x);\,\,\hat{P}f(x)=-i\hbar\pp_xf(x);\,\,[\hat{X},\hat{P}]=i\hbar$.
Then using $D_x^{\ga}$ there follows 
\bq\label{2.9}
\hat{x}=\left(\frac{\hbar}{mc}\right)^{1-\ga}x^{\ga};\,\,\hat{p}=-i\left(\frac{\hbar}{mc}\right)^{\ga}
mcD_x^{\ga}
\end{equation}
The classical and quantum Hamiltonians are now
\bq\label{2.10}
H_{class}=\frac{p^2}{2m}+\frac{1}{2}m\go^2x^2;\,\,H^{\ga}=\frac{\hat{p}^2}{2m}+\frac{1}{2}
m\go^2\hat{x}^2
\end{equation}
and the Schr\"odinger Hamiltonian becomes
\bq\label{2.11}
H^{\ga}\psi=\left[-\frac{1}{2m}\left(\frac{\hbar}{mc}\right)^2m^2 c^2D_x^{\ga}D_x^{\ga}
+\frac{1}{2}m\go^2\left(\frac{\hbar}{mc}\right)^{2(1-\ga)}x^{2\ga}\right]\psi=E\psi
\end{equation}
which is of the type (1.4) (with $D^{2\ga}=D^{\ga}D^{\ga}$).
The ``Hermiticity" of such an operator will depend of course on the choice of fractional
derivative and it can be shown that the Feller and Riesz fractional derivatives (but not
Caputo or Riemann-Liouville) will insure Hermiticity (cf. \cite{lask,hrmn}).  Here the Riesz
derivative is 
\bq\label{2.12}
D_R^{\ga}f(x)=\gG(1+\ga)\frac{Sin(\pi\ga/2)}{\pi}\int_0^{\infty}\frac{f(x+\xi)-2f(x)+f(x-\xi)}{\xi^{\ga+1}}d\xi;\,\,
(0<\ga<2)
\end{equation}
and the Feller derivative is 
\bq\label{2.13}
{}_FD_1^{\ga}f(x)=\gG(1+\ga)\frac{Cos(\pi\ga/2)}{\pi}\int_0^{\infty}\frac{f(x+\xi)-f(x-\xi)}{\xi^{\ga+1}}d\xi;\,\,(0\leq \ga<1)
\end{equation}
For a canonical
picture a scaled energy $E^q$ and coordinates can be introduced via
\bq\label{2.14}
\xi^{\ga}=\sqrt{\frac{m\go}{\hbar}}\left(\frac{\hbar}{mc}\right)^{1-\ga}x^{\ga};\,\,E=\hbar\go E^{\ga}
\end{equation} 
leading to the eigenvalues for $H^{\ga}$
\bq\label{2.15}
H^{\ga}\psi_n(\xi)=\frac{1}{2}\left[-D^{2\ga}_{\xi}+\xi^{2\ga}\right]\psi_n(\xi)=E'(n,\ga)\psi_n(\xi)
\end{equation}
Laskin (cf. \cite{lask}) has derived an approximate analytic solution within the framework
of the WKB approximation which has the advantage of being independent of the choice
of a specific definition of the fractional derivatives (cf. \cite{hrmn,lask} for more information on
this) and the result is
\bq\label{2.16}
E'(n,\ga)=\left[\frac{1}{2}+n\right]^{\ga}\pi^{\ga/2}\left[\frac{\ga\gG\left(\frac{1+\ga}{2\ga}\right)}{\gG(1/2\ga)}\right];\,\,n=0,1,2,\cdots
\end{equation}
(cf. also \cite{gwhn,lask,trsv,wgch}).  We write also for a harmonic oscillator (following \cite{hrmn})
the connection to q-deformation arises from $n>\sim x^{n\ga}$ with $(\bl\bl)\,\,E^q(n)=(\hbar\go/2)
([n]_q+[n+1]_q)$ while
\bq\label{2.17}
[n]_{\ga}|n>=\frac{\gG(1+n\ga)}{\gG(1+(n-1)\ga)}|n>;\,\,(\ga\sim q)
\end{equation}
for $n>0$ (as in (2.7)-(2.8)).  The q-deformation is then uniquely defined by $\ga$ via action
on pure states.
\\[3mm]\indent
{\bf REMARK 2.1.}
There is already however a huge literature on Fisher thermodynamics (TD), q-entropy, q-Fisher
information, non-extensive q statistics, multi-scale problems, etc., some of which we
looked at or referred to in \cite{crrr}.  More synthesis and organization is needed.  $\bs$

\section{Q DEFORMATION}
\renewcommand{\theequation}{3.\arabic{equation}}
\setcounter{equation}{0}

We have discussed fractals and fractional calculi in \cite{crrr} and refer also to \cite{ccgn,
clnd,cnsc,cgot}.  Some relations between fractional calculus and q-deformed calculus
were used but we want to expand upon the q-deformation framework here since the
relations to quantum groups etc. seem more ``intrinsic" in dealing with Schr\"odinger
equations (SE) and the quantum potential (QP) than simple recourse to fractional derivatives
(cf. Remark 3.1).
We will rely on \cite{bnds,c002,klsg} for background formulas and extract first a few
formulas from \cite{bnds}.  Thus one defines q-numbers $({\bf 3A})\,\,[x]=[(q^x-q^{-x})/(q-q^{-1})]
$ so $[1]=1,\,\,[2]=q+q^{-1}\,\,[3]=q^2+1+q^{-2}$, etc.
and notes that $({\bf 3B})\,\,e_q(az)=\sum_0^{\infty}(a^n/[n]!)z^n$.  Then
\bq\label{3.1}
D_x^qf(z)=\frac{f(qz)-f(q^{-1}z)}{(q-q^{-1})z}
\end{equation}
\bq\label{3.2}
D_x^q(az^n)=a[n]z^{n-1};\,\,D_z^qe_q(az)=ae_q(az)
\end{equation}
Further
\bq\label{3.3}
D_x^q[f(x)g(x)]=[D_x^qf(x)]g(q^{-1}x)+f(qx)[D_x^qg(x)]
\end{equation}
$$=(D_x^qg(x)f(q^{-1}x)+g(qx)[D_x^qf(x)]$$
\bq\label{3.4}
D_x^qf(x^n)=[n]x^{n-1}D_{x^n}^{q^n}(f(x^n);
\end{equation}
$$D_x^{q^n}f(x)=\frac{1}{[n]}\sum_0^{n-1}D_x^qf(q^{2k-(n-1)}x)$$
\bq\label{3.5}
\int_0^af(x)d_qx=a(q^{-1}-q)\sum_0^{\infty}q^{2n+1}f(q^{2n+1}a)
\end{equation}
\bq\label{3.6}
D_x^q\int f(x)d_qx=f(x)=\int D^q_xf(x)d_qx
\end{equation}

\indent
There are also Q numbers $({\bf 3C})\,\,[x]_Q=[(Q^x-1)/(Q-1)]$ with $({\bf 3D})\,\,
[x]=q^{1-x}[x]_Q$ when $Q=q^2$.  The exponential function is defined with a formula $({\bf 3E})\,\,
exp_Q(ax)=\sum_0^{\infty}[a^nx^n/[n]_Q!]$ with $e_Q(x)e_{1/Q}(-x)=1$.  The Q derivative is
\bq\label{3.7}
D_x^Qf(x)=\frac{f(Qx)-f(x)}{(Q-1)x}
\end{equation}
and 
\bq\label{3.8}
D_x^Qx^n=\frac{Q^nx^n-x^n}{(Q-1)x}=[n]_Qx^{n-1};\,\,[n]_Q=\frac{Q^n-1}{Q-1}
\end{equation}
while $({\bf 3F})\,\,D_x^Qe_Q(ax)=ae_Q(ax)$.  There are Leibnitz rules
\bq\label{3.9}
D^Q_x[f_1(x)f_2(x)]=[D^Q_xf_1(x)]f_2(Qx)+f_1(x)[D^Q_xf_2(x)];
\end{equation}
$$D^Q_x[(f_1(x)f_2(x)]=[(D^Q_x)f_1(x)]f_2(x)+f_1(Qx)[D^Q_xf_2(x)]$$
and for a quotient
\bq\label{3.10}
D_x^Q\frac{f_1(x)}{f_2(x)}=\frac{[D_x^Qf_1(x)]f_2(x)-f_1(x)[D^Q_xf_2(x)]}{f_2(Qx)f_2(x)}
\end{equation}
For the second derivative one has also
\bq\label{3.11}
[(D^Q_x)^2f(x)]=
\end{equation}
$$=(Q-1)^{-2}Q^{-1}x^{-2}[f(Q^2x)-(Q+1)f(Qx)+Qf(x)]$$
This leads to
\bq\label{3.12}
(D^Q_x)^nf(x)=
\end{equation}
$$=(Q-1)^{-n}Q^{-n(n-1)2}x^{-n}\sum_0^n\left[\begin{array}{c}
n\\
k
\end{array}\right]_Q(-1)^kQ^{k(k-1)/2}f(Q^{n-k}x)$$
Further
\bq\label{3.13}
\int_0^1f(x)d_Qx=(1-Q)\sum_0^{\infty}f(Q^k)Q^k
\end{equation}
It is also easily checked that $({\bf 3G})\,\,D^Q\int f(x)d_Qx=f(x)$.
We add to this a formula for $(D_x^q)^2$; thus
\bq\label{3.14}
D_x^q(D_x^qf(x))=D_x^q\left[\frac{f(qx)-f(q^{-1}x)}{(q-q^{-1})x}\right]=D_x^qh(x)=
\frac{h(qx)-h(q^{-1}x)}{(q-q^{-1})x}
\end{equation}
Consequently
\bq\label{3.15}
(D_x^q)^2f(x)=\frac{1}{(q-q^{-1})x}\left[\frac{f(q^2x)-f(x)}{(q-q^{-1})qx}-\frac{f(x)-f(q^{-2}x)}{(q-q^{-1})q^{-1}x}\right]=
\end{equation}
$$=\frac{1}{(q-q^{-1})x^2}\left[\frac{f(q^2x)}{q^2-1}+\frac{f(q^{-2}x)}{1-q^{-2}}-f(x)\left(\frac{1}
{(q^2-1)}+\frac{1}{(1-q^{-2})}\right)\right]$$
to compare with (3.11) for the Q derivative.
On the other hand we have from (3.4B) for $n=2$
\bq\label{3.16}
D^{q^2}_xf(x)=\frac{1}{q+q^{-1}}\left[\frac{f(x)-f(q^{-2}x)}{(q-q^{-1})x}+\frac{f(q^2x)-f(x)}{(q-q^{-1})x}\right]=
\end{equation}
$$=\frac{1}{(q^2-q^{-2})x}\left[f(q^2x)-f(q^{-2}x)\right]$$
\\[3mm]\indent
It is then appropriate to write down a SE for the q-derivative as
\bq\label{3.17}
i\hbar\psi_t=-\frac{\hbar^2}{2m}D^q_xD^q_x\psi +V\psi
\end{equation}
where $D^qD^q=(D^q)^2$ is specified in (3.14).  
Evidently $(D^q_x)^2\ne D^{2q}_x\ne D^{q^2}_x$.
\\[3mm]\indent
{\bf REMARK 3.1.}
Q calculus is a kind of time reversal of history, since $Qx\sim x+\gD x$ with $\gD x\sim (Q-1)x$.
On the other hand q-calculus seems to have a quantum meaning (cf. \cite{c002,klsg}).  Thus
one might expect a chain rule for Q calculus to follow from an equation
\bq\label{3.18}
D^Q_xf(x)=\frac{f(Qx)-f(x)}{(Q-1)x}=\frac{f(x+\gd x)-f(x)}{\gD x}
\end{equation}
with $Qx\sim x+\gD x$ and $\gD x\to 0$ when $Q\to 1$.  Then q-calculus simply works on
both sides of $q=1$.  This suggests working with $\psi=Rexp_Q(iS/\hbar)$ for the SE
provided one has a chain rule for $f(x)=exp_Q(u(x))$ with $\gD x\sim (Q-1)x$.  
According to \cite{hrmn} one can identify q-deformed derivatives with a fractional counterpart
based on the action of creation and annihilation operators on pure states via a formula
(in q-calculus) ($|n>\sim x^{n\ga}$)
\bq\label{3.19}
[n]_{\ga}|n>=\frac{\gG(1+n\ga)}{\gG(1+(n-1)\ga)}\,\,(n>0)
\end{equation} 
where $({\bf 3H})\,\,a|n>=\sqrt{[n]_q}|n-1>$ and $a^{\dg}|n>=\sqrt{[n+1]_q}|n+1>$
in the q-theory.
This suggests
that one should perhaps be able to write down a fractional counterpart.  We recall the
notation in Section 1 (used for MRL) where $d^{\ga}f=\ga!df$ and look at 
\bq\label{3.20}
D^Q_xu(x)=\frac{f(Qx)-f(x)}{(Q-1)x}=\frac{\gD f}{\gD x}\sim\frac{d^{\ga}f}{dx^{\ga}}
\end{equation}
and from \cite{jmre} there is a formula
\bq\label{3.21}
\frac{d^{\ga}u}{du^{\ga}}=(1-\ga)!u^{\ga-1}u^{1-\ga}\Rightarrow (u'_x(x))^{\ga}=
(1-\ga)!u^{\ga-1}u_x^{(\ga)}(x)
\end{equation}
However from \cite{kzch} there seems to be no general chain role for the Q-derivative, so
let us look for something else that might illuminate matters.  Thus consider for example
\bq\label{3.22}
\frac{f(Qu(Qx))-f(u(x))}{(Q-1)u(Q-1)x}=\frac{f(Qu(Qx))-f(u(Qx))+f(u,(Qx))-f(u(x))}{(Q-1)u(Q-1)x}=
\end{equation}
$$=\frac{D_u^Q(f(u(Qx))}{(Q-1)x}-\frac{D^Q_x(f(u(x))}{(Q-1)u}
$$
We will pick this up again in Section 4.  $\bs$

\section{DEFORMATIONS IN FRACTIONAL SPACES}
\renewcommand{\theequation}{4.\arabic{equation}}
\setcounter{equation}{0}

We have outlined briefly in \cite{crrr} some of the interaction of fractal space-time with fractional
calculus (cf. also \cite{ccgn,clnd,cnsc,cgot,eezn,ezek}) and the issue of fractional
calculus and QM is also developed in \cite{mtag,mtol}) in terms of a dimensionally deformed
D-calculus.  There are some similarities of this with the Q-deformed calculus discussed
above and we will survey some of this below.  The physics background for \cite{mtol} is
omitted here and we simply note that it deals with theory for an ``exciton" in a semi-conductor.
The theory in \cite{mtag}-1 involves a Bose-type oscillator in a fractional dimensional space and 
the mathematical aspects are summarized and polished in \cite{mtag}-2, which we examine
here.  Thus we sketch a new deformed calculus (D-deformed) beginning with a 1-dimensional
momentum operator $({\bf 4A})\,\,P=(1/i)(d/d\xi)$ where $\hbar$ is taken as 1.  In a fractional
space, sue to the inclusion of the integration weight 
\bq\label{4.1}
\frac{\gs(D)}{2}|\xi|^{D-1};\,\,\gs(D)=\frac{2\pi^{D/2}}{\gG(D/2)}
\end{equation}
the momentum operator is no longer Hermitian.  Working from the Wigner commutation relations for the canonical variables of a Bose-like oscillator one finds a momentum operator
\bq\label{4.2}
P=\frac{1}{i}\frac{d}{d\xi}+i\frac{(D-1)}{2\xi}R-i\frac{D-1}{2\xi}
\end{equation}
where R is a reflection operator.  This suggests a deformation of QM in a fractional space
via a new D-deformed derivative operator
\bq\label{4.3}
\frac{d_D}{d_D\xi}=\frac{d}{d\xi}+\frac{D-1}{2\xi}(1-R)
\end{equation}
with $({\bf 4B})\,\,P=(1/i)(d_D/d_D\xi)$.  Thus the deformed annihilation and creation operators
can be defined via
\bq\label{4.4}
a_D=\frac{1}{\sqrt{2}}\left(\xi+\frac{d_D}{d_D\xi}\right);\,\,a^{\dg}_D=\frac{1}{\sqrt{2}}\left(\xi
-\frac{d_D}{d_D\xi}\right)
\end{equation}
Then (cf. \cite{hrmn,mtag})
\bq\label{4.5}
a_D|2n>=\sqrt{2n}|2n-1>;\,\,a_D|2n+1>=\sqrt{2n+D}|2n>;
\end{equation}
$$a^{\dg}_D|2n>=\sqrt{2n+D}|2n+1>;\,\,a^{\dg}_D|2n+1>=\sqrt{2n+2}|2n+2>$$
Now, analogous to the q or Q factor, one introduces a D factor $({\bf 4C})\,\,[n]_D=n+[(D-1)/2][1-(-1)^n]$ so that (4.5) can be rewritten as $({\bf 4D})\,\,a_D|n>=\sqrt{[n]_D}|n-1>$ and $({\bf 4E})\,\,a^{\dg}_D|n>=\sqrt{[n+1]_D}|n+1>$.
Consequently one can define a D-deformed factorial as
\bq\label{4.6}
[n]_D=[n]_D[n-1]_D\cdots [1]_D[0]_D!=\left\{\begin{array}{cc}
\frac{2^n(n/2)\gG[(n+D)/2]}{\gG(D/2)} & n\,\,even\\
\frac{2^n[(n-1)/2]\gG[(n+d+1)/2]}{\gG(D/2)} & n\,\,odd
\end{array}\right.
\end{equation}
The eigenstates $|n>$ can be described via
\bq\label{4.7}
N_D|n>=n|n>;\,\,N_D=\frac{1}{2}\{a^{\dg}_D,a_D\}-\frac{D}{2};\,\,|n>=\frac{(a^{\dg}_D)^n}{\sqrt{[n]_D!}}0|>
\end{equation}
One can show directly that $(\bullet)\,\,a^{\dg}_Da_D=[n]_D;\,\,a_Da^{\dg}_D=[n+1]_D$
(cf. \cite{hrmn} for comparison with q and Q deformed calculi).  From the definitions
one can introduce now a D-deformed integration with
\bq\label{4.8}
f(\xi)=\int F(\xi)d_D\xi+C;\,\,\frac{d_Df(\xi)}{d_D\xi}=F(\xi)
\end{equation}
One expands upon this via
\bq\label{4.9}
\frac{d_Df(\xi)}{d_D(\xi)}=\left[1+\frac{D-1}{2\xi}(1-R)\int d\xi\right]\frac{df(\xi)}{d\xi}=F(\xi)\Rightarrow
\end{equation}
$$\Rightarrow \frac{df(\xi)}{d\xi}=\left[1+\frac{D-1}{2\xi}(1-R)\int d\xi\right]^{-1}F(\xi)$$
Consequently it follows that
\bq\label{4.10}
f(\xi)=\it F(\xi)d_D\xi=\sum_0^{\infty}\left[-\int dx\frac{D-1}{2\xi}(1-R)\right]^n\int d\xi F(\xi)\equiv
\end{equation}
$$\equiv \int F(\xi)d_D\xi=\sum_0^{\infty}(-1)^nI_n;\,\,I_{n+1}=\int\frac{D-1}{2\xi}(1-R)I_nd\xi;\,\,
I_0=\int F(\xi)d\xi$$
Then the following identities can be easily demonstrated
\bq\label{4.11}
\frac{d_D(f(\xi)g(\xi)}{d_D\xi}=g(\xi)\frac{d_Df(\xi)}{d_D\xi}+\frac{d_Dg(\xi)}{d_d\xi}Rf(\xi)+\frac{dg}{d\xi}(1-R)f(\xi)
\end{equation}
\bq\label{4.12}
\int g(\xi)\frac{d_Df(\xi)}{d_D\xi}d_D\xi=f(\xi)g(\xi)=\frac{d_Dg(\xi)}{d_D\xi}Rf(\xi)d_D\xi-
\int\frac{dg(\xi)}{d\xi}(1-R)f(\xi)d_D(\xi)
\end{equation}

\indent
The eigenstates can now be exhibited via
\bq\label{4.13}
P\Psi_p=-i\frac{d_D\psi_p}{d_D\xi}=p\Psi_p;\,\,\Psi_p=A_pE_D(ip\xi)
\end{equation}
where $E_D(x)$ is the D-deformed exponential function.  Thus
\bq\label{4.14}
\frac{d_D\xi^n}{d_D\xi}=[n]_D\xi^{n-1};\,\,\int \xi^nd_D\xi=\frac{\xi^{n+1}}{[n+1]_D}+const.
\end{equation}
\bq\label{4.15}
E_D(\xi)=\sum_0^{\infty}\frac{\xi^n}{[n]_D!};\,\,\frac{d_DE_D(\gl\xi)}{d_D\xi}=\gl E_D(\gl\xi);\,\,
\int E_D(\gl\xi)=\frac{E_D(\gl\xi)}{\gl}+c
\end{equation}

\indent
For eigenstates one has $({\bf 4F})\,\,a_D|\ga>=\ga|\ga>$ for coherent states where
\bq\label{4.16}
|\ga>=A_{\ga}\sum_0^{\infty}\frac{\ga^n}{\sqrt{[n]_D!}}|n>;\,\,A_{\ga}=\frac{1}{\sqrt{E_D(|\ga|^2)}};
\,\,|\ga>=\frac{E_D(\ga a^{\dg}]_D)|0>}{\sqrt{E_D(|\ga|^2)}}
\end{equation}
and we note that (4.7) is analogous to a formula in Q-deformation from \cite{bnds}, p. 547, with a Q deformed harmonic oscillator (with a background $q,\,q^{-1}$ oscillator satisfying
\bq\label{4.17}
a=q^{1/2}bq^{-N/2};\,\,a^{\dg}=q^{1/2}q^{-N/2}b^{\dg}
\end{equation}
\bq\label{4.18}
[N,a^{\dg}]=a^{\dg};\,\,[N,a]=-a;\,\,aa^{\dg}=-q^{\mp}a^{\dg}a=q^{\pm N}
\end{equation}
\bq\label{4.19}
|n>=\frac{(a^{\dg})^n}{\sqrt{[n]!}}|0>;\,\,N|n>=n|n>
\end{equation}
For the associated Q situation one has (cf. also \cite{arcn,kryn})
\bq\label{420}
[N,b^{\dg}]=b^{\dg};\,\,[N,b]=-b;\,\,b^{\dg}b=[N]_D;
\end{equation}
$$bb^{\dg}=[N+1]_Q;\,\,b|0>=0;\,\,
|n>=\frac{(b^{\dg})^n}{\sqrt{[n]_Q!}}|0>$$
which corresponds to a formula (4.7) with $D\sim Q$.
\\[3mm]\indent
{\bf REMARK 4.1.}
This seems to be a significant connection since it shows directly that a Q deformation can be
related to a dimensional factor D.  The connections between q-deformed situations and
fractional derivatives arising in \cite{hrmn} for example are somewhat more obscure with
a complicated formula relating q and the fractional term $\ga$.  It is perhaps suggested 
via $D\sim Q$ that one may begin with a fractional situation, with a deformation parameter D
(as in (4.1)-(4.2)), and proceed to a quantum type oscillator with $Q\sim D$ (as in
\cite{arcn,bnds,kryn}).  If so this would seem to be an important factor in modern work
on QM and gravity as in \cite{ccgn,clnd,cnsc,cgot}.  $\bs$

\section{DIFFERENTIAL EQUATIONS}
\renewcommand{\theequation}{5.\arabic{equation}}
\setcounter{equation}{0}

We want now to develop a SE type equation in the D and Q situations.  For a single degree
of freedom in a fractional space, following \cite{mtag}-1, one introduces a Cartesian type pseudo-coordinate $\xi\,\,(-\infty<\xi<\infty$).  The radial integration weight may be written as (cf. \cite{stil})
\bq\label{5.1}
\gs(D)r^{D-1}=\frac{\gs(D)}{2}|\xi|^{D-1};\,\,\gs(D)=\frac{2\pi^{D/2}}{\gG(D/2)}
\end{equation}
In this way the volume of the radius $R_0$ sphere in the fractional space is
\bq\label{5.2}
V(R,D)=\int_0^{R_0}\\gs(D)r^{D-1}dr=\int_{R_0}^{R_0}\frac{\gs(D)}{2}|\xi|^{D_1}d\xi=
\frac{\pi^{D/2}R_0)^D}{\gG(1+(D/2))}
\end{equation}
One takes $\hbar=1$ and the 1-dimensional momentum operator is then $({\bf 5A})\,\,
P=(1/i)(d/d\xi)$ which is not Hermitian for $D\ne 1$ and one also has to reject $({\bf 5B})\,\,
[\xi,P]=i$.  The most general wave-mechanical representation of P can be found by considering the general Wigner commutation relations for a Bose-like oscillator
\bq\label{5.3}
iP=[\xi,(P^2+\xi^2)/2];\,\,-i\xi=[P,(P^2+\xi^2)/2]
\end{equation}
The relations above can be rewritten as
\bq\label{5.4}
[P,S]=0;\,\,[\xi,S]=0;\,\,S=[\xi,P]-i
\end{equation}
This leads to the following expression
\bq\label{5.5}
<\xi'|S|\xi''>-=(\xi'-\xi)<\xi'|P|\xi''>-i\gd(\xi'-\xi'')
\end{equation}
On the other hand from (5.4)-B one has $({\bf 5C})\,\,<\xi'|S|\xi''>=2iA(\xi')\gd(\xi'+\xi'')$ where
A is a complex function.
This, with (5.5), shows that
\bq\label{5.6}
<\xi'|P|\xi''>=-i\gd'(\xi'-\xi'')+i\frac{A(\xi')}{\xi'}\gd(\xi'+\xi'')+B(\xi')\gd(\xi'-\xi'')
\end{equation}
Introducing the completeness condition $\int d\xi'|\xi'><\xi'|=1$ in (5.6) and writing the wave-function for a state $|\cdots)$ as $\psi(\xi')=<\xi',\cdots)$ it can be seen that
\bq\label{5.7}
P\psi(\xi')=-\frac{d\psi(\xi')}{d\xi'}+i\frac{A(\xi')}{\xi'}\psi(-\xi')+B(\xi')\psi(\xi')
\end{equation}
Thus the most general wave-mechanical representation for the momentum operator is
\bq\label{5.8}
P=\frac{1}{i}\frac{d}{d\xi}+i\frac{A(\xi)}{\xi}R+B(\xi)
\end{equation}
The S operator ({\bf 5C}) above can then be written as $({\bf 5D})\,\,S=2iA(\xi)R$.  From the anti-hermiticity $(S=-S^{\dg})$ and via substitution of (5.8) and ({\bf 5D}) in (5.4) one arrives
at restrictions on A and B of the form
\bq\label{5.9}
A^*(\xi)=A(-\xi);\,\,\frac{dA(\xi)}{d\xi}+i[B(\xi)+B(-\xi)]A(\xi)=0
\end{equation}
For a 1-dimensional fractional-dimensional momentum one requires the hermiticity of P
(cf. \cite{mtag}-3) and via (5.8) and (5.1) B must be given by $({\bf 5E})\,\,B(\xi)=i[(D-1)/2\xi]$.
Then ({\bf 5E}) and (5.4) lead to $A(\xi)=constant$ and one can write
\bq\label{5.10}
\na=\frac{d}{d\xi}-\frac{A}{\xi}R+\frac{D-1}{2\xi}
\end{equation}
Then requiring $\na u=0\iff u=constant$ one gets $({\bf 5F})\,\,A=[(D-1)/2]$ and 
\bq\label{5.11}
P=\frac{1}{i}\frac{d}{d\xi}+i\frac{(D-1)}{2\xi}R-i\frac{(D-1)}{2\xi}=\frac{1}{i}\frac{d}{d\xi}+i\frac{(D-1)(R-1)}{2\xi}
\end{equation}
This is essentially a Dunkle operator (cf. \cite{dnkl}) and R is the reflection operator which
reverses sign of the argument (i.e. $R=\pm 1$, even or odd); thus
\bq\label{5.12}
\psi^{even}(\xi)=\psi(\xi)+\psi(-\xi)=(1+R)\psi(\xi);\,\,\psi^{odd}(\xi)=\psi(\xi)-\psi(-\xi)=(1-R)\psi(\xi)
\end{equation}
Note that in a 1-dimensional space the
Heisenberg uncertainty becomes $\sqrt{(\gD\xi)^2(\gD P)^2}\geq 1/2$ and this follows from
({\bf 5B}).  However in a fractional dimensional space one obtains
\bq\label{5.13}
\sqrt{((\gD\xi)^2)}\sqrt{(\gD p)^2)}\geq
\left\{\begin{array}{cc}
D/2 & even\,\, states\\
(2-D)/2 & odd\,\, states\\
1/2 & otherwise
\end{array}\right.
\end{equation}
(we refer to \cite{mtag}-1 for details).
\\[3mm]\indent
Now consider a free particle in a fractional-dimensional space and write the Hamiltonian as
\bq\label{5.14}
H=\frac{P^2}{2}=-\frac{1}{2}\left[\frac{d^2}{d\xi^2}+\frac{(D-1)}{\xi}\frac{d}{d\xi}-
\frac{(D-1)(1-R)}{2\xi^2}\right]
\end{equation}
(assuming $\hbar=m=1$).  The SE equation here is stationary in the form $H\psi=E\psi$ and the eigenfunctions
satisfy
\bq\label{5.15}
\left[\frac{d^2}{d\xi^2}+\frac{(D-1)}{\xi}\frac{d}{d\xi}+2E\right]\psi^{even}(\xi)=0;
\end{equation}
$$\left[\frac{d^2}{d\xi^2}+\frac{(D-1)}{\xi}\frac{d}{d\xi}-\frac{(D-1)}{\xi^2}+2E\right]\psi^{odd}(\xi)=0$$
based on $R=\pm 1$.  This leads to
\bq\label{5.16}
\psi_p(\xi)=A_p(|p\xi |^{1-(D/2)}[J_{(D/2)-1}(|p\xi |)+
isgn(p\xi)J_{D/2}(|p\xi |)];\,\,
A_p=\sqrt{\frac{|p|^{D-1}}{2\gs(D)}}
\end{equation}
Here $sgn(x)= 1\,\, (x>0)$ and $=-1\,\,(x<0)$ and the $J_{\nu}(x)$ are Bessel functions.
Note from \cite{klst} (p. 32) that Bessel functions of the first kind have an expression
\bq\label{5.17}
J_{\nu}(z)=\sum_0^{\infty}\frac{(-1)^k(z/2)^{2k+\nu}}{k!\gG(\nu+k+1)}
\end{equation}
An interesting picture of the behavior of a free particle in a fractional dimensional space may now be obtained by computing the position dependence of the probability density
\bq\label{5.18}
\rho_p=\frac{\gs(D)}{2}|\xi|^{D-1}|\psi_p|^2
\end{equation}
(cf. \cite{mtag}-1 for discussion).
\\[3mm]\indent
For the study of a fractional dimensional Bose type oscillator one puts (5.11) in $H=(P^2+\xi^2)/2$ to obtain the Hamiltonian (cf. (5.14))
\bq\label{5.19}
H=-\frac{1}{2}\left[\frac{d^2}{d\xi^2}+\frac{(D-1)}{\xi}\frac{d}{d\xi}-\frac{(D-1)-(D-1)R}{2\xi^2}-
\xi^2\right]
\end{equation}
Certain classes of integrable many-body systems (Calogero models) have been analyzed
using the Dunkl operator leading to a Hamiltonian essentially of the form (5.19).  This
suggests a connection between the N-body Calogero problem and the fractional dimensional
Bose-like oscillator (cf. \cite{mtag}-1 for more discussion).

\section{ANOTHER KIND OF QP}
\renewcommand{\theequation}{6.\arabic{equation}}
\setcounter{equation}{0}

Given a SE based on (5.13) one might look for solutions $\psi=rexp(i{\mf S}/\hbar)$
or $\psi=rE_D(i\mf S/\hbar)$ (as in $(\bullet)$ of Section 1) where $E_D$ is the D-deformed exponential as in
(4.14)-(4.15) (for $\hbar=m=1$).  We reserve R here for the reflection operator below.
This also may not be realistic but we are motivated 
to be heuristic here.  The $t$ variable can be omitted in looking for the QP (quantum potential) and
we can simply use (5.11) to express H in terms of ${\mc D}_D=d_D/d^D\xi=
{\mc D}_{\xi}+(\ga_D/\xi)$ with $\ga_D=(D-1)(1-R)/2$ (cf. (4.3)).  Thus, following \cite{mtag}-1,
\bq\label{6.1}
H=-\frac{1}{2}\left[\left({\mc D}_D-\frac{\ga_D}{\xi}\right)^2+\frac{(D-1)}{\xi}\left({\mc D}_D
-\frac{\ga_D}{\xi}\right)-\frac{\ga_D}{\xi^2}-\xi^2\right]=
\end{equation}
$$=-\frac{1}{2}\left[{\mc D}_D^2-\frac{2\ga_D}{\xi}{\mc D}_D+\frac{\ga_D}{\xi^2}{\mc D}_D+
\right.$$
$$\left.+\frac{\ga_D^2}{\xi^2}+\frac{(D-1)}{\xi}{\mc D}_D-\frac{(D-1)\ga_D}{\xi^2}-\frac{\ga_D}{\xi^2}-\xi^2\right]=$$
$$=-\frac{1}{2}\left[{\mc D}_D^2-\frac{2\ga_D}{\xi}{\mc D}_D+
\frac{\ga_D}{\xi^2}{\mc D}_D+\frac{\ga_D^2}{\xi^2}
+\frac{(D-1){\mc D}_D}{\xi}-\frac{D\ga_D}{\xi^2}-\xi^2\right]=$$
$$=-\frac{1}{2}\left[{\mc D}^2_D+\left(\frac{D-1-2\ga_D}{\xi}+\frac{\ga_D}{\xi^2}\right){\mc D}_D+
\left(\frac{\ga_D^2-D\ga_D}{\xi^2}\right)-\xi^2\right]$$
We note also that $({\bf 6A})\,\,\ga_D(\ga_D-D)=(1/2)[-D(1+R)+(R-1)]$ and $({\bf 6B})\,\,
2\ga_D=(D-1)(1-R)$.
\\[3mm]\indent
Now working with the $\xi$ variable as in (5.19) we can examine a stationary SE, $H\psi=E\psi$ with $\psi=rexp(iS/\hbar),\,\,S=S(\xi)$, and $\hbar=1$ in order to generate a ``putative"
QP.  Thus 
\bq\label{6.2}
\psi_{\xi}=r_{\xi}exp(\cdot) +ir_{\xi}S_{\xi}exp(\cdot)
\end{equation}
$$\psi_{\xi\xi}=r_{\xi\xi}exp(\cdot)+2ir_{\xi}S_{\xi}exp(\cdot)
+irS_{\xi\xi}exp(\cdot)+r(iS_{\xi})^2exp(\cdot)$$
The real term gives then  from (5.19) (cf. also (5.14))
\bq\label{6.3}
(H\psi)_{real}=E\psi\Rightarrow Er\sim-\frac{1}{2}\left[r_{\xi\xi}-rS_{\xi}^2+\frac{D-1}{\xi}r_{\xi}-
\frac{\ga_Dr}{\xi^2}-\xi^2r\right]
\end{equation}
For even (resp. odd) eigenfunctions in (5.15) one has $R=1\sim\ga_D=0$ or $R=-1\sim\ga_D=D-1$.  Thus for odd eigenfunctions (6.3) becomes (cf. (5.15))
\bq\label{6.4}
H\psi_{real}= E\psi_{real}\Rightarrow Er\sim -\frac{1}{2}\left[r_{\xi\xi}-rS_{\xi}^2+\frac{D-1}{\xi}r_{\xi}-\frac{(D-1)r}{\xi^2}-\xi^2r\right]
\end{equation}
and the kinetic energy term involves $(1/2)(S_{\xi}^2$).  The $\xi^2r$ term is already a potential energy so this leaves a residual energy term
\bq\label{6.5}
{\mf Q}\sim -\frac{r_{\xi\xi}}{2r}-\frac{D-1}{2r\xi}r_{\xi}+\frac{D-1}{2\xi^2}
\end{equation}
which can perhaps play the role of a quantum potential (QP).  Note also for $R=-1$ and
$\ga_D=D-1$ with ${\mc D}_D=d_{\xi}+[(D-1)/\xi]$ one has
\bq\label{6.6}
{\mc D}_D^2=d_{\xi}^2+\frac{D-1}{\xi}d_{\xi}-\frac{D-1}{\xi^2}+\frac{(D-1)^2}{\xi^2}=
\end{equation}
$$=d_{\xi}^2+\frac{D-1}{\xi}d_{\xi}+\frac{1}{\xi^2}(D-1)(2-D)$$
Hence
\bq\label{6.7}
\frac{{\mc D}_D^2r}{2r}=-{\mf Q}-\frac{(D-1)(1-2D)}{2\xi^2}
\end{equation}
which might be of some interest. 
The formula (6.7) does remind one of the ``standard"
formulas for a QP of the form $({\bf 6C})\,\,\gb(\gD r/r)$ or $\gag(D^{2\ga}r/r)$.  For
completeness we list also the stationary state formula for the even eigenfunctions (where $\ga_D=0$ and ${\mc D}_D=d_{\xi}$)
\bq\label{6.8}
H\psi_{real}=E\psi_{real}\sim -\frac{1}{2}\left[r_{\xi\xi}-rS_{\xi}^2+\frac{D-1}{\xi}r_{\xi}-\xi^2r\right]
\end{equation}
Here we encounter the relation $({\bf 6D})\,\,{\mf Q}\sim -[{\mc D}_D^2r/2r]$ more in the traditional mold.
\\[3mm]\indent
At this point
it would be nice to relate the operators ${\mc D}_D$ and $D_Q$ from (3.7) and (4.3) (cf. also 
(4.16)-(4.20) for Q and D connections).  
It is interesting that the powers of $\xi$ or $x$
which arise in fractional and deformed calculus seem to have a natural physical role.
For relations between q-deformation and generalized statistics see e.g. \cite{crrr,lgsw,
lsns} and references there; see also \cite{bayn,c002,crrr,klsg,masi,mabu,wess,zhng}
for q or Q deformed quantum mechanics.


\newpage
\begin{thebibliography}{cccc}






%
\bibitem{arcn} M. Arik and D. Coon, Jour. Math. Phys., 17 (1976), 524
%
\bibitem{bayn} S. Bayin, math-ph 1203.4556 and 1208.1142
%
\bibitem{bnds} D. Bonatsos and C. Daskaloyanis, Prog. Part. Nuclear Phys., 43 (1999), 537-618
%
\bibitem{ccgn} G. Calcagni, hep-th 0912.3142, 1001.0571, 1012.1244, 1106.0295, 1204.2550, and 1205.5046
%
\bibitem{clnd} G. Calcagni and G. Nardelli, math-ph 1202.5383; hep-th 1004.5144
%
\bibitem{cnsc} G. Calcagni, G. Nardelli, and M. Scalisi, hep-th 1207.4473
%
\bibitem{cgot} G. Calcagni, S. Gielen, and D. Oriti, gr-qc 1201.4151
%
\bibitem{cpct} A. Carpinteri and P. Cornetti, Chaos, Solitons, and Fractals, 13 (2002), 85-94
%
\bibitem{c002} R. Carroll, Calculus revisited, Kluwer, 2002
%
\bibitem{c067} R. Carroll, Fluctuations, information, gravity, and the quantum
potential, Springer, 2006;  On the quantum potential, Arima
Publ., 2007
%
\bibitem{c009} R. Carroll, On the emergence theme of physics,
World Scientific, 2010
%
\bibitem{crar} R. Carroll, Quantum Potential as Information: A mathematical survey,  in:
New trends in quantum information, Aracne Edit.,
2010, pp. 155-189
%
\bibitem{cola} R. Carroll, 
Progress in Physics, 1 (2012), 27-29 (gr-qc 1110.3059) and 2 (2012), 82-86
%
\bibitem{crlr} R. Carroll, Quantum theory, deformations, and integrability, Elsevier, NorthHolland, 2000
%
\bibitem{clrr} R. Carroll, math-ph 1007.4744, gr-qc 1010.1832 and 1104.0383
%
\bibitem{crrr} R. Carroll, Some topics in thermodynamics and quantum mechanics,
physics 1211.1898
%
\bibitem{cruz} W. da Cruz, cond-mat 9811007
%
\bibitem{dnkl} C. Dunkl, Trans. Amer. Math. Soc., 311 (1989), 167
%
\bibitem{erly} R. Earnshaw and E. Riley (editors), Brownian motion ..., Nova Publ., 2011, pp. 1-69
%
\bibitem{eezn} A. Ezan, Phys. Lett. A, 225 (1997), 235-238
%
\bibitem{ezek} A. Erzan and J. Eckmann, Phys. Rev. Lett., 78 (1997),3245-3248
%
\bibitem{ezgb} A. Erzan and A. Gorbon, cond-mat 98078052
%
\bibitem{fred} B. Frieden, Physics from Fisher information,
Cambridge Univ. Press, 1998; Science from Fisher information,
Springer, 2004
%
\bibitem{frgy} B. Frieden and R. Gatenby, Exploratory data analysis using Fisher information, Springer, 2007
%
\bibitem{gwhn} C. Godinho, J. Weberszpil, and J. Helayei-Neto, CSF, 45 (2012 (765-771
%
\bibitem{gros} G. Gr\"ossing, 
Entropy, 12 (2010), 1975-2044;  quant-ph 
0201035, 0204070, 0205048, 0311109, 0404030, 0410236, 0508079, 0711.4954, 0806.4462, 0808.3539, and 0812.3561;  Phys. Lett. A, 372 (2008), 4556-4562; Found.
Phys. Lett., 17 (2004), 343-362
%
\bibitem{hrmn} R. Herrmann, math-ph 0510099; physics 0711.3701, 0805.3434, and 1007.1084; math.GM 1107.4205;
Fractional calculus, World Scientific, 2011; Fraktionale Infinitesimalrechnung, Altenstadt, 2008
%
\bibitem{jmre} G. Jumarie, Applied Math. Letters, 18 (2005), 739-748 and 817-826, 22 (2009), 378-385 and
1659-1664; Chaos, Solitons,
and Fractals, 32 (2007), 969-987; 40 (2009),1428-1448 and 41 (2009), 1590-1604; 12 (2001), 2577-2587;
Acta Math. Sinica, 28 (2012), 1741-1768
%
\bibitem{kzch} V. Kac and P. Cheung, Quantum calculus, Springer, 2002
%
\bibitem{klst} A. Kilbas, H. Srivastava, and J. Trujillo, Theory and applications of fractional differential
equations, North-Holland (Elsevier), 2006
%
\bibitem{klsg} A. Klimyk and K. Schm\"udgen, Quantum groups and their representation, Springer, 1997
%
\\bibitem{kblv} V. Kobelev, math-ph 1292.274; Chaos, 16 (2006), 043117
%
\bibitem{kryn} V. Kuryshkin, Ann. Fond. L. de Broglie, 51 (1980), 111
%
\bibitem{lask} N. Laskin, math-ph 1009.5533 and 0811.1769; Phys. Rev. E, 66 (2002),
056108
%
\bibitem{lgsw} A. Lavagno and P. Narayana Swamy, cond-mat 0911.1635
%
\bibitem{lsns} A. Lavagno, A. Scarfone, and P. Narayana Swamy, cond-mat 0509477 and
0504748; Rept. Math. Phys., 55 (2005), 423-433
%
\bibitem{masi} M. Masi, cond-mat 0611300
%
\bibitem{mtag} A. Matos-Abiague, Jour. Phys. A, 34 (2001), 3125-3138; quant-ph 0107062;
Physica Scripta, 62 (2000), 106
%
\bibitem{mtol} A. Matos-Abiague, L. Oliveira, and M. de Dios-Leyva, Phys. Rev. B, 58
(1998), 4072-4076
%
\bibitem{mabu} S. Muslih, O. Agrawal, and D. Baleanu, Int. Jour. Theor. Physics, 49 (2010),
1746-1752
%
\bibitem{nabr} M. Naber, math-ph 0410028
%
\bibitem{stil} F. Stillinger, Jour. Math. Phys., 18 (1977), 1224-1234
%
\bibitem{strz} R. Strichartz, Differential equations on fractals,
Princeton Univ. Press, 2006
%
\bibitem{trsv} V. Tarasov, physics 0602096, 0602208, 0604491, 0804.0586, 0907.2699,1107.4205, 1107.5682, and 1107.5749
%
\bibitem{wgch} J. Weberszpil, C. Godinho, A. Cherman, and J. Helayei-Neto, math-ph 1206.2513
%
\bibitem{wess} J. Wess, math-ph 9910013; hep-th 0408080 and 0607251; Nucl. Phys. B,
Supplement 18B (1990), 302-312
%
\bibitem{zhng} J. Zhang, hep-th 0310043
%




\end{thebibliography}
\end{document}